\documentclass[aps,prl,amsmath,amssymb,floatfix,lengthcheck,reprint]{revtex4-1}

\usepackage{graphicx}

\begin{document}
\title{RFSQUID-Mediated Coherent Tunable Coupling Between a Superconducting Phase Qubit and a Lumped Element Resonator}
\author{M.S. Allman, F. Altomare, J.D. Whittaker, K. Cicak, D. Li, A. Sirois, J. Strong, J.D. Teufel, R.W. Simmonds}
\email{simmonds@boulder.nist.gov}
\address{National Institute of Standards and Technology, 325 Broadway, Boulder, Colorado 80305-3328, USA}
\date{\today}

\begin{abstract}
We demonstrate coherent tunable coupling between a superconducting phase qubit and a lumped element resonator.  The coupling strength is mediated by a flux-biased RF SQUID operated in the non-hysteretic regime.  By tuning the applied flux bias to the RF SQUID we change the effective mutual inductance, and thus the coupling energy, between the phase qubit and resonator .  We verify the modulation of coupling strength from $0$ to $100\;MHz$ by observing modulation in the size of the splitting in the phase qubit's spectroscopy, as well as coherently by observing modulation in the vacuum Rabi oscillation frequency when on resonance.  The measured spectroscopic splittings and vacuum Rabi oscillations agree well with theoretical predictions.

\end{abstract}

\maketitle
Superconducting qubit research has made tremendous strides in recent years.  Superconducting qubits are routinely made with coherence lifetimes approaching $1\;\mu{s}$ and beyond \cite{Clarke}.  Also, a number of coupled qubit experiments with fixed coupling between qubits have been performed \cite{Majer, Sillanpaa, TsaiA, TsaiB, WellstoodA, WellstoodB, MartinisA, MartinisB}.  Any real superconducting quantum computer, however, will be composed of an intricate network of many qubits coupled to each other in various ways, as well as coherent ``quantum buses'' that will manage the shuttling of quantum information between distant qubits.  This means that it will become increasingly difficult to implement quantum information processing between many coupled quantum circuit elements with fixed coupling between elements.  The need to control the coupling between various elements, such as qubit-qubit interactions or qubit-quantum bus interactions is essential.  A number of ways of implementing tunable coupling between quantum circuit elements have been proposed in recent years \cite{NiskanenB, Averin, Plourde, Wallquist, Brink} and performed experimentally \cite{Fay,Niskanen, Hime, Ploeg, Harris}.  One rather conceptually simple way of implementing tunable coupling, proposed by \cite{Brink}, involves use of a flux-biased RF-SQUID, operated in the non-hysteretic regime, as a tunable ``flux-transformer'' between elements.  We have employed such a coupling scheme to coherently couple a superconducting phase qubit to a lumped element resonator.

\begin{figure}[!htbp]
\centering
\includegraphics[width=\columnwidth]{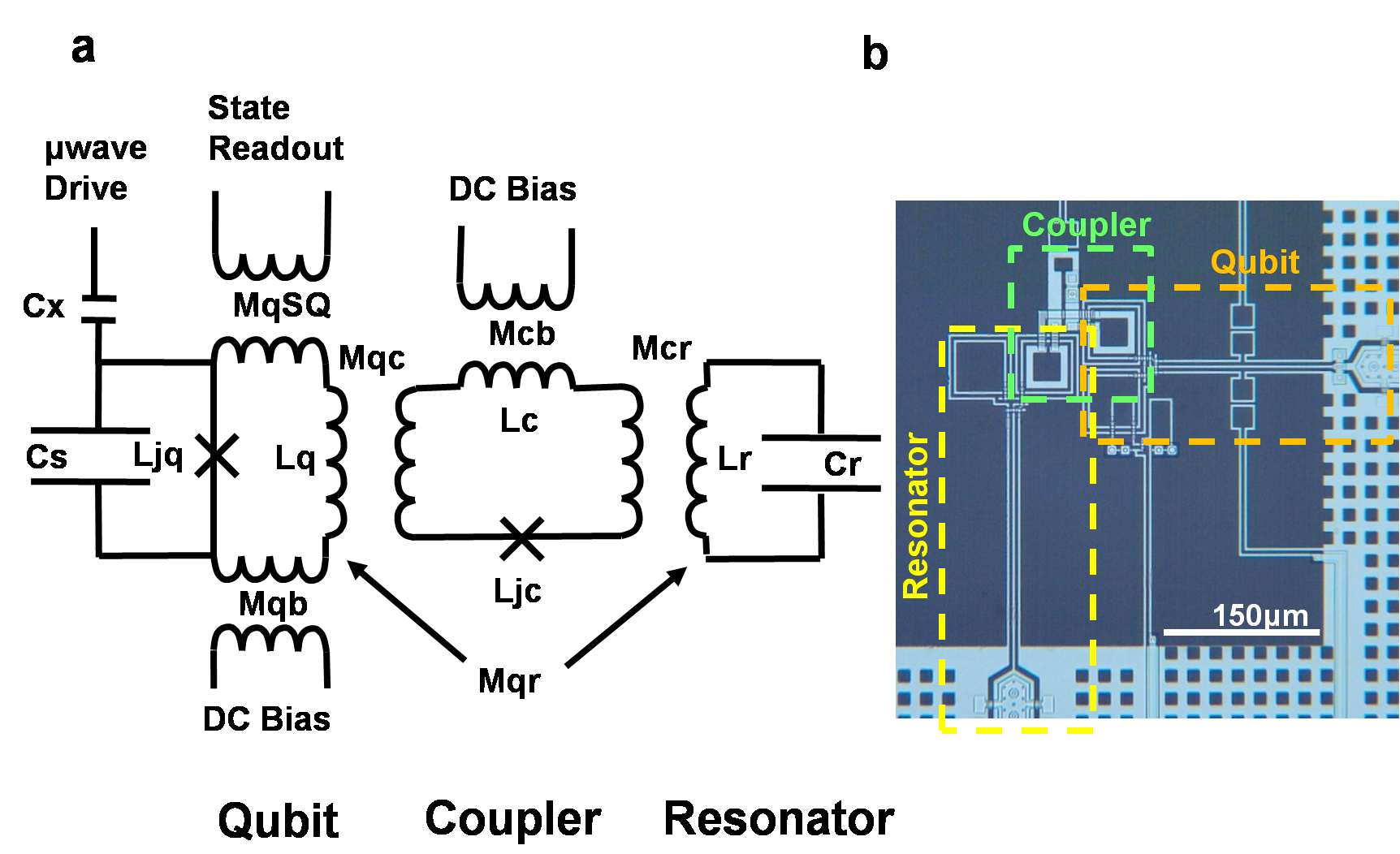}
\caption{(a) Circuit diagram for the phase qubit, coupler and resonator.  The qubit parameters are $I_{q0}\simeq0.6\;\mu{A}$, $C_{qs}\simeq0.6\;pF$, $L_q\simeq1000\;pH$, $\beta_{q}\simeq1.8$, and $M_{qc}\simeq60\;pH$.  The coupler parameters are $I_{c0}\simeq0.9\;\mu{A}$, $L_c\simeq200\;pH$, $C_{jc}\simeq0.3\;pF$ and $\beta_c\simeq0.5$. The resonator parameters are $L_r\simeq1000\;pH$, $C_r\simeq0.4\;pF$, and $M_{cr}\simeq60\;pH$. (b) Optical micrograph of the circuit.} \label{Device}
\end{figure}

The circuit for this experiment is illustrated in Figure \ref{Device}.  It is composed of a phase qubit, with critical current $I_{q0}$, shunt capacitance $C_{qs}$, and geometric inductance $L_q$, coupled through a mutual inductance $M_{qc}$, to the RF SQUID, referred to as ``the coupler''.  The coupler has a critical current $I_{c0}$, geometric inductance $L_c$, and junction capacitance $C_{jc}$.  It is coupled through a mutual inductance $M_{cr}$, to the lumped element resonator of geometric inductance $L_r$ and capacitance $C_r$.  All the junctions are via-style ion-mill junctions, and the capacitors were fabricated by use of ``vacuum'' capacitor technology \cite{Cicak}.  There is also a residual mutual inductance $M_{qr}$ between the qubit and resonator, which was gradiometrically designed to be as small as possible.

The phase qubit is also coupled to external control and readout circuitry.  A dc bias line, coupled to the qubit loop via a mutual inductance $M_{qb}$, provides an external flux bias to the qubit.  This bias controls the nonlinear Josephson inductance of the qubit that controls the energy level spacing between qubit states as well as level anharmonicity.  The qubit is operated in a flux bias regime that creates an approximately cubic metastable potential well of sufficient anharmonicity to reliably isolate the lowest two metastable states of the well \cite{Simmonds}.

A microwave drive capacitively coupled via series capacitance $C_x$ provides the excitation energy to drive transitions between the two lowest qubit levels, labeled $\vert{g}\rangle$ and $\vert{e}\rangle$, respectively.  A short ($\sim5\;ns$) measure pulse is then applied to induce tunneling of the $\vert{e}\rangle$ state to the adjacent stable well.  The state of the qubit is read out via a DC SQUID coupled to the qubit's geometric inductance via a mutual inductance $M_{qSQ}$\cite{Cooper}.

The qubit's circulating current, $I_q$ couples an amount of flux $M_{qc}I_q$ into the coupler, generating a circulating current $I_c$ governed by the relation

\begin{equation}
i_c=-\sin\left(2\pi\phi_{x} +\beta_ci_c\right), \label{EQNCurrent}
\end{equation}
where $i_c = I_c/I_{c0}$ is the normalized circulating coupler current, $\phi_{x}=\left(\Phi_{x}+M_{qc}I_q\right)/\Phi_0$ is the net external flux applied to the coupler, and $\beta_c=2{\pi}L_cI_{c0}/\Phi_{0}<1$.  This current then couples flux to the resonator via mutual inductance $M_{cr}$.  For a given change in qubit flux, the flux change seen by the resonator depends on the offset current circulating in the coupler due to the external bias flux, $\Phi_x$.  The result is a tunable effective mutual inductance between the qubit and resonator given by

\begin{eqnarray}
M_{eff}(\Phi_{x})&=&M_{qc}M_{cr}\frac{I_{c0}}{\Phi_0}\frac{\partial i_c}{\partial \phi}_{x}\nonumber\\
&=&\frac{M_{qc}M_{cr}}{L_c}\frac{\beta_c \cos \left[2\pi\phi_{x} +\beta_ci_c\right]}{\left( 1+\beta_c\cos %
\left[2\pi\phi_x +\beta_ci_c\right] \right)}.
\label{eq:Meff}
\end{eqnarray}

From equation \ref{eq:Meff} we see that the effective mutual inductance can be tuned anywhere between the following extrema:
\begin{eqnarray*}
\left( M_{eff}\right) _{\max } &=&\frac{M_{qc}M_{cr}}{L}\frac{\beta_c}{1-\beta_c}\ \ for\ \ n_{odd}\\
\left( M_{eff}\right) _{\min } &=&-\frac{M_{qc}M_{cr}}{L}\frac{\beta_c}{1+\beta_c}\ \ for\ \ n_{even},
\end{eqnarray*}
by choosing $\Phi_{x}$ such that
\begin{equation}
\Phi_{x} = n\frac{\Phi_0}{2} - M_{qc}I_q,
\end{equation}
corresponding to a null circulating current in the coupler.
In particular, $M_{eff}=0$, when the coupler circulating current is at the critical current.  It is also worth noting that in the limit that $\beta_c\rightarrow 1$, $(M_{eff})_{max}$ increases without bound.

\begin{figure}[!htbp]
\centering
\includegraphics[width=\columnwidth]{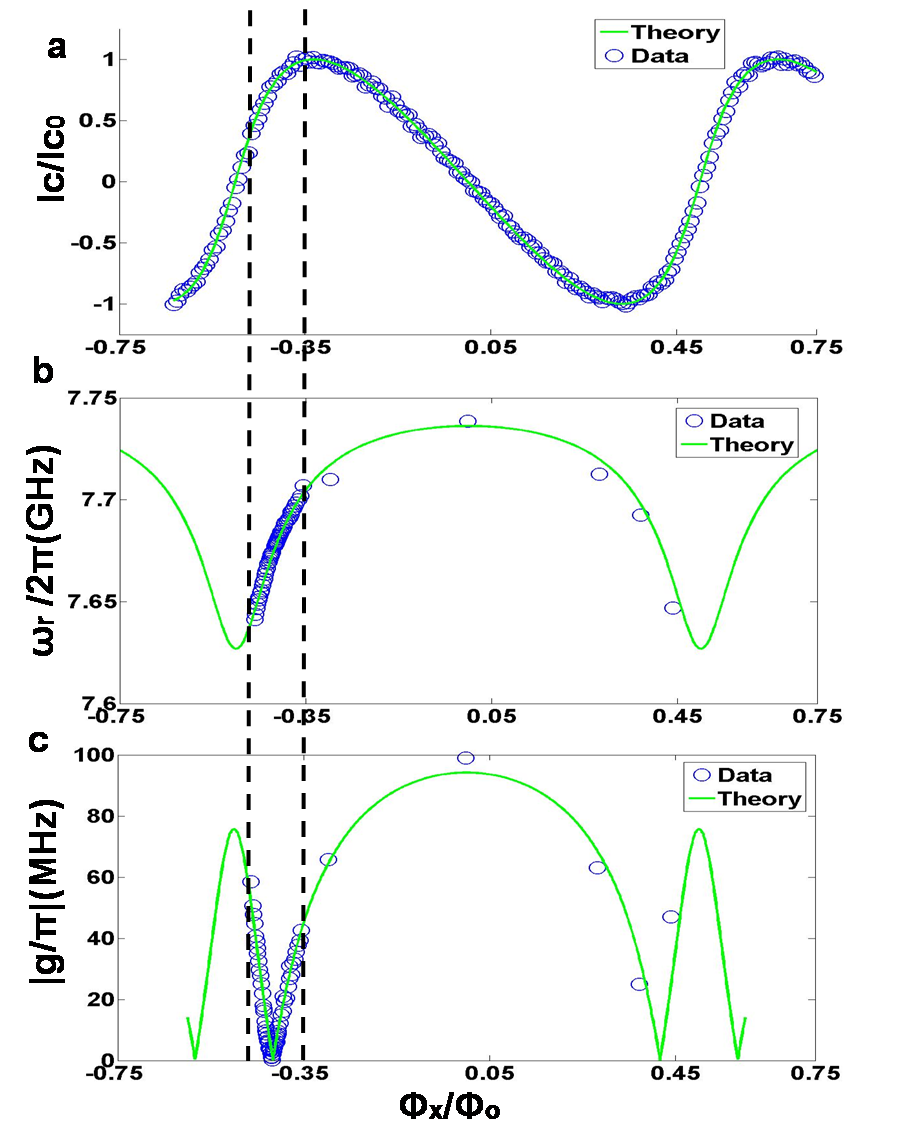}
\caption{Measurements of the dependence of $I_c$, $\omega_r$, and $g_c$ on applied coupler flux, $\Phi_x/\Phi_0$.  The vertical dashed lines bracket the applied flux ranges for the waterfall data shown in Figure\ref{CombinedWFall}. (a) The measured circulating coupler current as a function of applied coupler flux along with the theoretical fit giving $\beta_c=0.51$. (b) Measured resonator frequency as a function of applied coupler flux, along with theoretical fit using $\beta_c$ extracted from (a).  The fit yields $\omega_{r0}/2\pi=7.710\;GHz$. (c) Measured coupling strength as a function of applied coupler flux along with the theoretical fit using parameters extracted from the theory fits in (a) and (b).} \label{IGFCombined}
\end{figure}

An interesting consequence of the changing effective mutual inductance between the qubit and resonator is that the resonator's frequency modulates with the applied flux as
\begin{equation}
\omega_r(\Phi_x) = \omega_{r0}\sqrt{1+\frac{M_{cr}}{L_rM_{qc}}M_{eff}(\Phi_x)}
\end{equation}
where $\omega_{r0} = {1}/{\sqrt{L_rC_r}}$.  The measured resonator frequency is shown in Figure \ref{IGFCombined}(b).

We approximate the Hamiltonian of our system using the Jaynes-Cummings model in the rotating-wave approximation,

\begin{eqnarray}
\hat{H}=\hat{H}_{q} +\hat{H}_r +\hat{H}_{I}(\Phi_x)+\hat{H}_{\kappa}+\hat{H}_{\gamma},
\label{eq:Hamiltonian}
\end{eqnarray}
where $\hat{H}_{q} = \frac{1}{2}\hbar \omega _{q}\hat{\sigma}_{qz}$ is the qubit Hamiltonian, $\hat{H}_r = \left( \hat{a}_{r}\hat{a}_{r}^{\dag }+\frac{1}{2}\right) \hbar \omega _{r}$ is the resonator Hamiltonian, and the interaction term, $\hat{H}_{I}(\Phi_x) = \hbar g_{c}(\Phi_{x})\left(\hat{\sigma}^{+}_{q}\hat{a}_{r}+\hat{\sigma}^{-}_{q}\hat{a}_{r}^{\dag }\right)$ describes the exchange of a single excitation between the qubit and resonator at a rate proportional to
\begin{equation}
g_{c}(\Phi_x)\approx\frac{\omega_r}{2}\frac{M_{total}(\Phi_x)}{\sqrt{L_qL_r}},
\label{eq:gc}
\end{equation}
where $M_{total}(\Phi_x) = M_{eff}(\Phi_x)+M_{qr}$ incorporates the direct mutual inductance between the qubit and resonator.  The last two terms $H_{\kappa}$ and $H_{\gamma}$, describe the coupling of the resonator and qubit to environments that give rise to the resonator decay rate $\kappa$, and qubit decay rate $\gamma$ \cite{Blais}.

The lowest two levels of a the qubit and resonator form a four-dimensional joint Hilber space.  We label the qubit's ground and first excited states as $\vert{g}\rangle$ and $\vert{e}\rangle$, respectively, and the resonator's ground and first excited states as $\vert0\rangle$ and $\vert1\rangle$, respectively.  According to equation \ref{eq:Hamiltonian}, when the qubit is on resonance with the resonator, so that the detuning $\Delta=\omega_q-\omega_r=0$, individual eigenstates of the qubit and resonator, given by $\vert{g0}\rangle, \vert{e0}\rangle, \vert{g1}\rangle$, and $\vert{e1}\rangle$ are no longer the eigenstates of the coupled system.  The new eigenstates are found to be $\vert{g0}\rangle$, and $\vert{e1}\rangle$ and the symmetric and antisymmetric superpositions $\vert{\pm}\rangle=\left(\vert{g1}\rangle\pm\vert{e0}\rangle\right)/\sqrt{2}$.  The corresponding energy eigenvalues are $E_{g0}=0, E_{e1}=2\hbar\omega_r$ and $E_{\pm}=\hbar(\omega_r\pm{g_c(\Phi_x)})$.

The coupler is first calibrated by sweeping its external flux bias, $\Phi_x$, and measuring the effect on the tunneling probability of the $\vert{g}\rangle$ state of the qubit.  By tracking the required applied qubit flux $\Phi_q$, to maintain a constant total qubit flux $\phi_q=\left(\Phi_q+M_{qc}I_c\right)/\Phi_0$ such that the $\vert{g}\rangle$ state tunneling probability is approximately $10\%$, we can determine the circulating current in the coupler as a function of $\Phi_x$.  Figure \ref{IGFCombined}(a) shows the measured coupler circulating current as a function of applied coupler bias flux.

\begin{figure}[!htbp]
\centering
\includegraphics[width=\columnwidth]{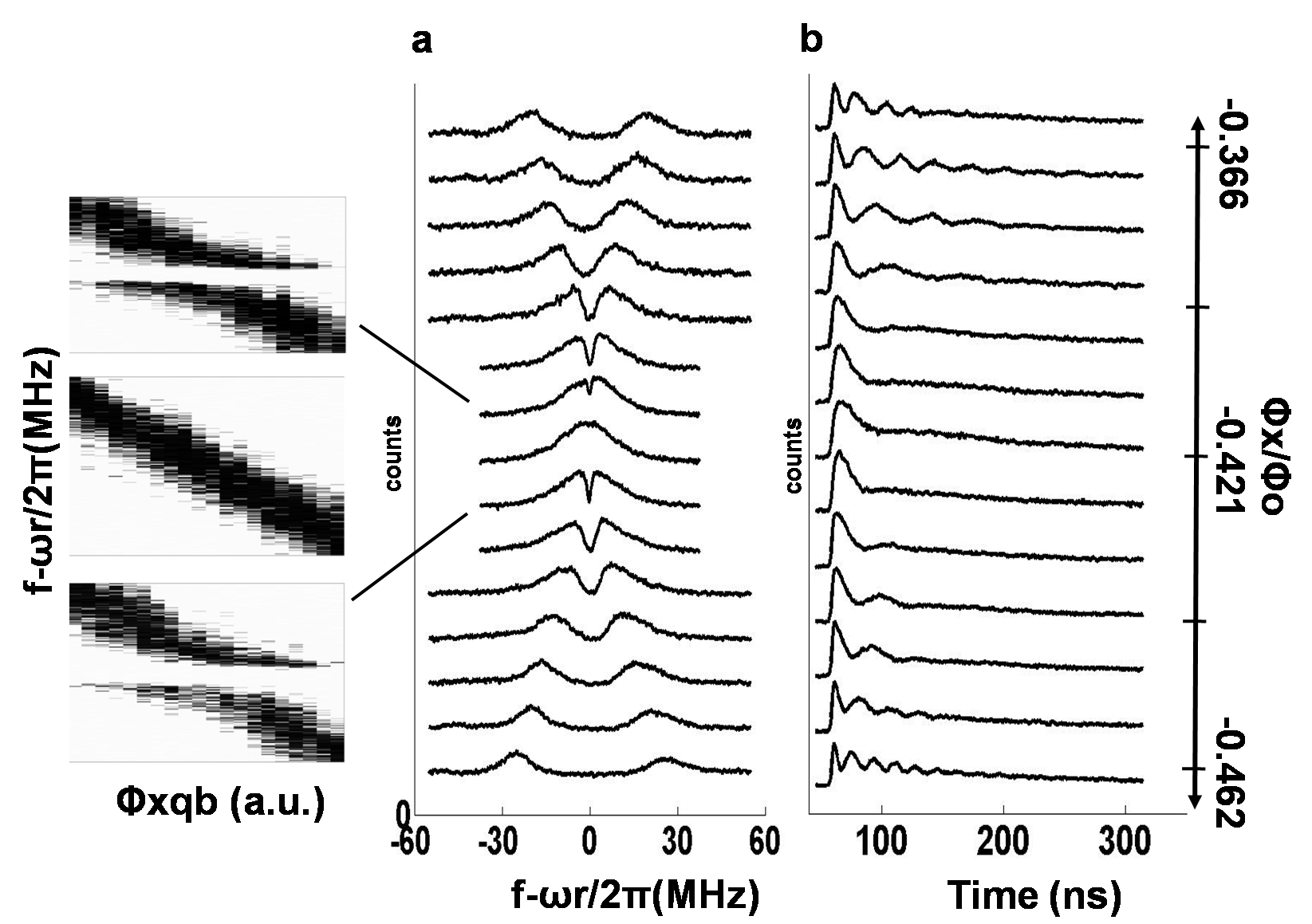}
\caption{Spectroscopic and time-domain data over the range $\Phi_x/\Phi_0=-0.462$ to ${\Phi_x}/{\Phi_0}=-0.366$ bounded by the vertical dashed lines in Figure \ref{IGFCombined}.  (a) Waterfall plot of the spectroscopic measurements of the $\vert{\pm}\rangle$ states showing the splitting transition from $g_c(-0.462)/\pi\simeq50\;MHz$ through $g_c(-0.421)/\pi=0$ to $g_c(-0.366)/\pi\simeq40\;MHz$.  The inset to the left is a 3D plot of the qubit spectroscopy showing the avoided crossing transition through zero for applied coupler flux values close to $\Phi_x=-0.421$.  (b) The corresponding vacuum Rabi measurements demonstrating coherent modulation in the coupling strength $g_c(\Phi_x)$.} \label{CombinedWFall}
\end{figure}

The next step in the experiment is to demonstrate the effect of the coupler on the quantum mechanical interactions between the qubit and cavity.  We first look for a cavity interaction using well-established spectroscopic techniques \cite{Simmonds, Cooper}.  By use of figure \ref{IGFCombined}(a) the coupler is set to the desired coupling strength and then qubit spectroscopic measurements are performed.  When the qubit transition frequency nears the resonant frequency of the resonator, an avoided crossing occurs, splitting the resonant peak into two peaks.  When the qubit frequency exactly matches the resonator's frequency ($\Delta=0$) the size of the spectroscopic splitting is minimized to $g(\Phi_x)/\pi$.  This whole cycle is repeated for different flux biases applied to the coupler.  We observe the size of the zero-detuning splitting modulate from a maximum of $g_{max}/\pi\approx100\;MHz$ down to no splitting (Figure\ref{CombinedWFall} (a)).  The spectroscopic measurements are a good indicator that the coupler is working, but we do not consider them to be proof of coherent coupling between the qubit and resonator, because the length of the microwave pulse is longer ($\simeq500\;ns$) than the lifetime of the qubit.

To demonstrate coherent tunable coupling we measure the vacuum Rabi oscillation period between the qubit and resonator as a function of applied coupler flux.  According to equation \ref{eq:Hamiltonian}, neglecting dissipation, the zero-detuning probability of finding the system in the state $\vert{e}0\rangle$  as a function of time is periodic and given by

\begin{eqnarray}
\vert\left\langle e0\right\vert \left\vert \Psi \left( t\right) \right\rangle\vert^2 &=& \vert\langle{e0}\vert{e^{i\hat{H}t/\hbar}}\vert{e0}\rangle\vert^2\nonumber\\
&=&\frac{1}{2}\left(1+\cos \left( 2g_c(\Phi_x)t\right)\right).
\label{eq:probe0}
\end{eqnarray}

When dissipation is added the oscillatory behavior of equation \ref{eq:probe0} decays exponentially with a rate given by $\gamma_{avg} = (\gamma+\kappa)/2$. When $\vert{4g(\Phi_x)}\vert=\vert{\kappa-\gamma}\vert$, the system is critically damped and the $\vert{e0}\rangle$ state decays at the rate $\gamma_{avg}$ with no oscillations \cite{Haroche}.  The qubit's decay rate was measured to be $\gamma=1/T_1=1/135\;ns$.  The resonator's decay rate was not measured in this experiment directly, but previous experiments have found them to be smaller than $\kappa\sim1/1000\;ns$ \cite{Cicak}.  If we take $\gamma>>\kappa$, then the critical coupling strength where the oscillations are expected to disappear is $g_c/\pi\simeq1\;MHz$.

\begin{figure}[!htbp]
\centering
\includegraphics[width=\columnwidth]{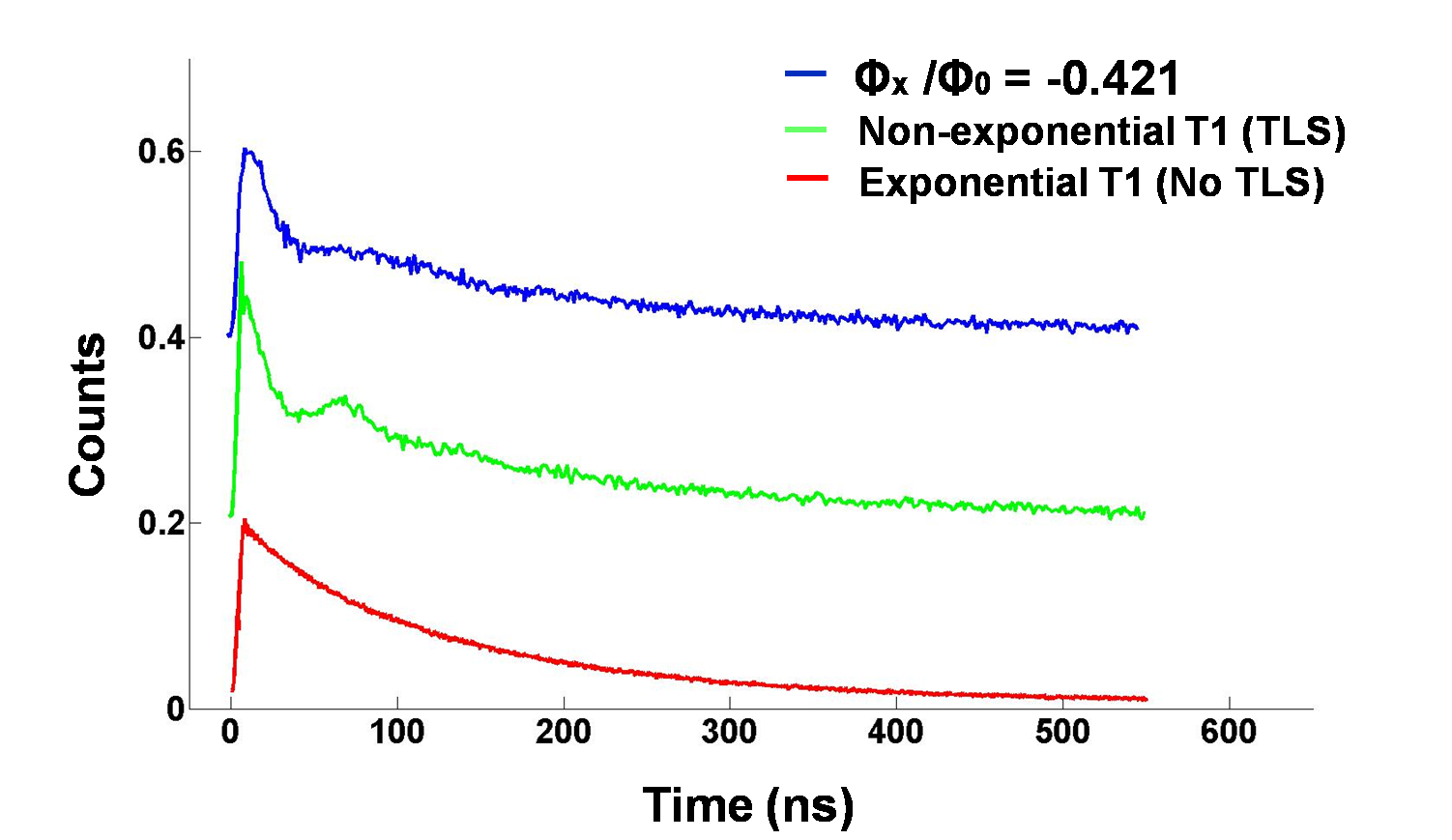}
\caption{A higher resolution trace of the occupation probability of the $\vert{e0}\rangle$ state when $\Phi_x/\Phi_0=-0.421$ along with exponential $T_1$ and non-exponential $T_1$ measurements taken at a qubit frequencies largely detuned from the resonator.  The non-exponential $T1$ trace showed no evidence of a TLS interaction in the corresponding spectroscopy.} \label{T1}
\end{figure}

Experimentally, we excite the $\vert{e0}\rangle$ state by applying a short $(\tau_p\simeq5-10\;ns)$ pulse with the qubit on resonance with the resonator.  The pulse is fast enough that the resonator remains in its ground state during state preparation.  We then measure the state of the qubit as a function of time.  Figures \ref{IGFCombined}(b,c) and \ref{CombinedWFall} summarize the spectroscopic and time domain measurements.  For $g(\Phi_x)/\pi>10\;MHz$, the vacuum Rabi data are used to determine the coupling strength by applying a Fast Fourier Transform (FFT) to the measured probability data.  For $g(\Phi_x)/\pi<10\;MHz$, the FFT method is less reliable and the coupling strength is determined from the size of the splitting in the spectroscopy data at zero detuning.  We clearly see the splitting in the spectroscopy shrink to zero when the coupler is ``off'', but the corresponding time-domain data do not appear to be exponential, as predicted by eq \ref{eq:Hamiltonian} when $g_c(\Phi_x)=0$.  There appears to be a rapid drop followed by a slow $(\simeq7\;MHz)$ oscillation in the data (Figure \ref{T1}).

There is a higher order coupling channel not included in equation \ref{eq:gc}, resulting from the finite but small direct mutual inductance between the qubit and resonator, $M_{qr}$.  This residual coupling strength is given by\cite{Ashab}

\begin{equation}
g_{residual} \approx\frac{M_{qr}}{\sqrt{L_qL_r}}\frac{M_{qc}}{\sqrt{L_qL_c}}\frac{M_{cr}}{\sqrt{L_cL_r}}\frac{\omega_r^4}{\omega_c^3}.
\end{equation}

For our design parameters $g_{residual}\sim10kHz$, much too weak to account for the residual effect seen in the data.  We believe the residual coupling effect is due to weakly coupled, spurious two-level system fluctuators (TLSs) interacting with the qubit at this frequency \cite{Simmonds}.  We have used a scan of vacuum Rabi data that confirms these types of weak oscillations throughout the entire spectroscopic range, even at frequencies far detuned from the resonator.  This indicates interactions with weakly coupled TLSs not seen in traditional spectroscopy measurements.  Figure \ref{T1} compares the vacuum Rabi data taken at $\Phi_x/\Phi_0=-0.421$ and the exponential and non-exponential $T_1$ data taken at qubit frequencies far detuned from the resonator and where no TLS splittings were visible in the spectroscopic data.

We have demonstrated coherent tunable coupling between a superconducting phase qubit and a lumped element resonator, using a separate, flux-biased RFSQUID as a mediating element.  Spectroscopically, the coupling strength was observed to modulate from a maximum $100\;MHz$ to zero.  The Vacuum Rabi oscillation frequency was observed to agree well with the spectroscopic measurements for $\vert{g_c(\Phi_x)/\pi}\vert\geq7\;MHz$.  The residual oscillations for weaker coupling strengths are believed to be due to spurious TLSs in the junction barrier and not the result of a residual coupling effect from the coupler.

\bibliographystyle{apsrev4-1}

%

\end{document}